\documentclass[prc,twocolumn,epsfig,nofootinbib,floatfix,showpacs,superscriptaddress]{revtex4-2}
\usepackage{graphics}

\usepackage{epsfig}
\usepackage{amsfonts}
\usepackage{amsmath}
\usepackage{float}
\usepackage{bm}
\usepackage{mathrsfs}
\usepackage{makecell}
\usepackage{siunitx}
\usepackage{color}
\usepackage{scrextend}
\usepackage{tablefootnote}
\usepackage[breaklinks=true,colorlinks=true,linkcolor=blue,urlcolor=blue,citecolor=red]{hyperref}

\begin{document}

\title{Wavelet analysis of monopole  strength  in highly deformed  $^{24}$Mg}

    \author{A.~Bahini}
\email[]{bahini@lpccaen.in2p3.fr}
\affiliation{School of Physics, University of the Witwatersrand, Johannesburg 2050, South Africa}
\affiliation{iThemba Laboratory for Accelerator Based Sciences, Somerset West 7129, South Africa}
\affiliation{Université de Caen Normandie, ENSICAEN, CNRS/IN2P3, LPC Caen UMR6534, F-14000 Caen, France}
\author{V.~O.~Nesterenko}
\email[]{nester@theor.jinr.ru}
\affiliation{Laboratory of Theoretical Physics, Joint Institute for Nuclear Research, Dubna, Moscow Region 141980, Russia}
\affiliation{Dubna State University, Dubna, Moscow Region 141980, Russia}
\author{P.~von~Neumann-Cosel}
\affiliation{Institut f\"{u}r Kernphysik, Technische Universit\"{a}t Darmstadt, D-64289 Darmstadt, Germany}
\author{P.-G.~Reinhard}
\affiliation{Institute for Theoretical Physics II, University of Erlangen, D-91058 Erlangen, Germany}
\author{J.~Carter}
\affiliation{School of Physics, University of the Witwatersrand, Johannesburg 2050, South Africa}	
\author{N.~A.~Ashurko}
\affiliation{National Research Tomsk State University, Tomsk 634050, Russia}	
\affiliation{National Research Tomsk Polytechnic University, Tomsk 634050, Russia}
\author{R.~Neveling}
\affiliation{iThemba Laboratory for Accelerator Based Sciences, Somerset West 7129, South Africa}	
\author{A.~Repko}
\affiliation{Institute of Physics, Slovak Academy of Sciences, 84511 Bratislava , Slovakia}
\author{I.~T.~Usman}
\affiliation{School of Physics, University of the Witwatersrand, Johannesburg 2050, South Africa}	

\begin{abstract}

\noindent \textbf{Background:}
Experimental data on $\alpha$-particle inelastic scattering for monopole excitations in $^{24}$Mg 
in the excitation-energy region $E_{\rm x} = 9$--$25$~MeV, obtained at the iThemba Laboratory for Accelerator Based Sciences (iThemba LABS), have been analyzed within a fully self-consistent quasiparticle random-phase approximation (QRPA) framework using two Skyrme parametrizations.~A good overall agreement with the experimental data is achieved, particularly with the SkP$^{\delta}$ force, which corresponds to a low nuclear incompressibility of $K_{\infty}=202$~MeV.

\noindent \textbf{Objective:}
To extract energy scales, by means of wavelet analysis, characterizing the observed fine structure of the isoscalar giant monopole resonance (ISGMR) as well as the low-energy region $10$$-$$18$ MeV of the deformation-induced monopole-quadrupole coupling (MQC) in order to investigate the damping mechanism contributing to their decay widths.

\noindent \textbf{Methods:}
Characteristic energy scales are extracted from the fine structure using continuous wavelet transforms.~The experimental results are compared to QRPA calculations employing the Skyrme parameterizations SkP$^{\delta}$ and SVbas.

\noindent \textbf{Results:}
A significant, if not decisive, impact of the MQC strength on the wavelet power spectra is observed across the entire excitation-energy range of $10$--$24$~MeV.~Wavelet features derived from the QRPA and from unperturbed two-quasiparticle (2qp) monopole strengths are compared.~The results demonstrate that the residual interaction plays a key role in reproducing realistic wavelet powers and characteristic energy scales.~Overall, a continuous range of scales $\delta E$$=$$200$$-$$1000$~keV is obtained rather than distinct isolated scales.~The deformation softness of $^{24}$Mg is found to significantly influence both the monopole strength distribution and the wavelet characteristics.

\noindent \textbf{Conclusions:}
Similar to the case of the isovector giant dipole resonance (IVGDR), the dominant role of Landau damping is confirmed as the main decay mechanism of the ISGMR in the highly deformed nucleus $^{24}$Mg.~At the same time, the remaining discrepancies between experiment and theory highlight the possible importance of couplings to more complex configurations.

\end{abstract}

\maketitle
\date{\today}

\section{Introduction}
\label{intr}

For several decades, the isoscalar giant monopole resonance (ISGMR) has remained a subject of intensive experimental and theoretical investigation~\cite{Harakeh,Garg}.~This collective mode is of particular importance as it provides direct information on the nuclear incompressibility~\cite{Blaizot_PR80}.~In deformed nuclei, however, the ISGMR exhibits an additional remarkable feature — the deformation-induced coupling between the ISGMR and the $\beta$-vibrational $L K$$=$$20$ branch (here, $K$ denotes the projection of the total angular momentum $L$ onto the symmetry axis of an axially deformed nucleus~\cite{bohr}) of the isoscalar quadrupole giant resonance (ISGQR)~\cite{Kva_PRC16,Colo_PLB20}.~In well-deformed systems, this coupling can become very strong, leading to a pronounced splitting of the ISGMR into two distinct components: the main ISGMR part and a lower-energy fraction arising from the monopole–quadrupole coupling (MQC)~\cite{Kva_PRC16}.

An exceptionally strong MQC is expected in highly deformed light nuclei such as $^{24}$Mg, where the prolate axial quadrupole deformation reaches a value of $\beta$$\sim$$0.6$~\cite{Garg_PLB15,Garg_PRC16,Kva_24Mg_EPJ_WC16}.~Both the main ISGMR and its MQC component were recently observed in $^{24}$Mg through inelastic $\alpha$-particle scattering experiments performed at the iThemba Laboratory for Accelerator Based Sciences (iThemba LABS).~The results were analyzed within the self-consistent Skyrme quasiparticle random-phase approximation (QRPA) framework~\cite{Adsley_PRC21,Bahini_PRC22}.~The analysis of Ref.~\cite{Bahini_PRC22} indicates that, due to the large deformation splitting, the main ISGMR in $^{24}$Mg extends over a wide excitation-energy interval, starting approximately at $16$~MeV and appears rather fragmented.~In contrast, the MQC manifests itself as a structure dominated by a narrow peak at $13.9$~MeV.~The characteristics of this peak are expected to be sensitive to the properties of both the ISGMR and the ISGQR.~Since the ISGQR energy depends on the isoscalar effective mass $m_0^*/m$~\cite{Adsley_PRC21,Nest_IJMPE08}, the MQC features should be influenced by both the nuclear incompressibility and the effective mass (see also the discussion on the coupling between these two quantities in nuclear matter~\cite{Chabanat}).~Altogether, the main ISGMR and its MQC component have different physical origins, each requiring separate consideration.

In recent years, wavelet analysis~\cite{Shev_PRC08,Shev_PRC09} has been widely employed to investigate the fine structure of various giant resonances (GRs); see, for example, studies of the ISGQR in both spherical~\cite{Shev_PRC08,Shev_PRC09} and deformed nuclei~\cite{Kureba_PLB18}, analyses of the isovector giant dipole resonance (IVGDR) in {\it sd}-shell nuclei~\cite{fearick2018}, in $^{40,48}$Ca~\cite{carter2022}, and in heavy spherical and deformed nuclei~\cite{donaldson2020fine,poltoratska2014}, as well as investigations of the ISGMR in spherical systems~\cite{Bahini_PRC24,Bahini_NPA2025}.~Wavelet analysis provides valuable access to information on the widths of GRs and, consequently, on their main decay properties~\cite{Peter2019review}.~It is well established that the total width of a GR ($\Gamma_{\rm GR}$) can be approximately decomposed into three components~\cite{Harakeh,Goeke82}:\\
\begin{equation}
\Gamma_{\rm GR} = \Delta\Gamma + \Gamma^{\uparrow} + \Gamma^{\downarrow}~,
\end{equation}\\
where $\Delta\Gamma$ denotes the Landau damping (fragmentation of the collective mode into nearby one-particle one-hole ($1$p-$1$h) excitations), $\Gamma^{\uparrow}$ represents the escape width associated with particle emission, and $\Gamma^{\downarrow}$ is the spreading width arising from the coupling of $1$p-$1$h states to more complex configurations, including two-particle two-hole ($2$p-$2$h) and many-particle many-hole (np-nh) states.~Following previous studies~\cite{Shev_PRC09,Bahini_PRC24}, the damping of the resonance through the spreading width $\Gamma^{\downarrow}$ plays a dominant role in spherical nuclei for the ISGQR and ISGMR, while for the IVGDR, Landau damping remains significant even in heavy systems such as $^{208}$Pb~\cite{Peter2019review}.~In deformed nuclei, on the other hand, the widths of giant resonances are mainly governed by Landau damping~\cite{Donald_PLB18,Kureba_PLB18}, at least in the case of the ISGMR~\cite{Kva_PRC16}.~Therefore, QRPA calculations restricted to one-phonon configurations are generally sufficient to describe the centroid energy and total width of the ISGMR.~Nevertheless, it remains an open question whether QRPA alone can adequately account for the {\it fine} structure of the ISGMR, particularly in highly-deformed light nuclei, where GRs often exhibit pronounced structural fragmentation~\cite{Harakeh,Garg}.

The QRPA-based wavelet analysis provides additional valuable insight into the dominant damping mechanisms shaping the ISGMR structure, particularly the role of Landau damping.~In the present work, we perform a wavelet analysis of the isoscalar monopole (IS$0$) excitations in the highly-deformed nucleus $^{24}$Mg.~On the experimental side, we employ the recent $\alpha$-particle inelastic scattering data obtained at iThemba LABS.~The theoretical calculations are carried out within the fully self-consistent QRPA framework~\cite{Repko_arXiv15,Repko_EPJA17_pairing,Kvasil_EPJA19} using the Skyrme parametrizations SkP$^{\delta}$~\cite{SkPd} and SVbas~\cite{SVbas}.~The IS$0$ strength distributions in the ISGMR and MQC energy regions are analyzed separately.~To elucidate the role of Landau damping, the wavelet spectra obtained from the QRPA calculations are compared with those derived from the unperturbed two-quasiparticle (2qp) configurations.~In addition, the impact of the deformation softness of $^{24}$Mg on the wavelet results is inspected.

The paper is organized as follows.~Section~\ref{sec2} provides a brief description of the experimental setup and the data.~The details of the QRPA calculations and the corresponding results for the ground state of $^{24}$Mg are outlined in Sec.~\ref{sec3}.~The monopole-quadrupole coupling (MQC) effects in $^{24}$Mg are illustrated and discussed in Sec.~\ref{sec4}.~The fundamentals of the wavelet analysis method are summarized in Sec.~\ref{sec5}.~The main results and their discussion are presented in Sec.~\ref{sec6}, and the conclusions are drawn in Sec.~\ref{sec7}.

\section{Experimental details}
\label{sec2}
The details of the experimental procedure followed in this study are given in Ref.~\cite{Bahini_PRC22}.~As such, only the main points are summarized here.~The experiment was performed at the Separated Sector Cyclotron (SSC) facility of iThemba LABS.~A beam of $196$ MeV $\alpha$-particles was inelastically scattered off a self-supporting $^{24}$Mg target with an areal density of $0.23$~mg/cm$^2$.~The reaction products were momentum analyzed using the K$600$ magnetic spectrometer positioned at laboratory scattering angles $0^{\circ}$ and $4^{\circ}$.~The isoscalar monopole (IS$0$) strength distributions were extracted using the Difference-of-Spectra (DoS) technique \cite{DoSpaper}.~An energy resolution of approximately $70$ keV full width at half-maximum (FWHM) was achieved, revealing significant fine structure in the isoscalar monopole strength distributions in $^{24}$Mg.

\section{Model and calculation details} 
\label{sec3}
The calculations are performed within the matrix QRPA model~\cite{Repko_arXiv15,Repko_EPJA17_pairing,Kvasil_EPJA19} based on the Skyrme functional \cite{Ben_RMP2003}.~The model is fully self-consistent in that:
(i) both the mean field and the residual interaction are derived from the same Skyrme functional;
(ii) the contributions of all time-even densities and time-odd currents from the functional are included; (iii) both the particle-hole and pairing-induced particle-particle channels are taken into account; and (iv) the Coulomb direct and exchange parts are included consistently in both the mean field and the residual interaction.

The Skyrme parameterizations SkP$^{\delta}$~\cite{SkPd} and SVbas~\cite{SVbas} were employed in the present calculations.~As summarized in Table~\ref{tab1:forces}, these forces exhibit markedly different nuclear incompressibilities $K_{\infty}$, while yielding comparable isoscalar effective masses $m_0^*/m$.~The SkP$^{\delta}$ interaction employs a volume-type pairing, whereas SVbas uses a density-dependent surface pairing.~Both parameterizations have been successfully applied in previous studies of the ISGMR and the deformation-induced monopole-quadrupole coupling (MQC) in deformed nuclei such as $^{24}$Mg and $^{28}$Si~\cite{Bahini_PRC22}, as well as in deformed Mo isotopes~\cite{Colo_PLB20}.~It has been shown~\cite{Colo_PLB20,Bahini_PRC22} that SkP$^{\delta}$, with its relatively low incompressibility $K_{\infty}$$=$$202$~MeV, provides a significantly better description of the experimental data than SVbas, which has a higher incompressibility $K_{\infty}$$=$$234$~MeV.

\begin{table}[t]  
\caption{Main properties of the Skyrme parameterizations SkP$^{\delta}$ and SVbas used in the QRPA calculations.}
\begin{center}
\setlength{\arrayrulewidth}{0.5pt}
\setlength{\tabcolsep}{0.4cm}
\renewcommand{\arraystretch}{1.4}	
\begin{tabular}{lccc}
\hline\hline
Force & $K_{\infty}$ (MeV) & $m_0^*/m$ & Pairing type \\
\hline
SkP$^{\delta}$ & 202 & 1.00 & Volume \\
SVbas & 234 & 0.90 & Surface \\
\hline\hline
\end{tabular}
\label{tab1:forces}
\end{center}
\end{table}

\begin{table}[t]  
\caption{Experimental~\cite{nndc} and calculated quadrupole deformation parameters $\beta$ and binding energies per nucleon $BE/A$ for $^{24}$Mg.}
\begin{center}
\setlength{\arrayrulewidth}{0.5pt}
\setlength{\tabcolsep}{0.6cm}
\renewcommand{\arraystretch}{1.4}	
\begin{tabular}{lcc}
\hline\hline
 & $\beta$ & $BE/A$ (MeV) \\
\hline
Experiment & 0.613 & 8.26 \\
SkP$^{\delta}$ & 0.545 & 7.82 \\
SVbas & 0.527 & 8.02 \\
\hline\hline
\end{tabular}
\label{tab2:def+BE}
\end{center}
\end{table}

\begin{figure} 
 \begin{center}
  \includegraphics[scale=0.32]{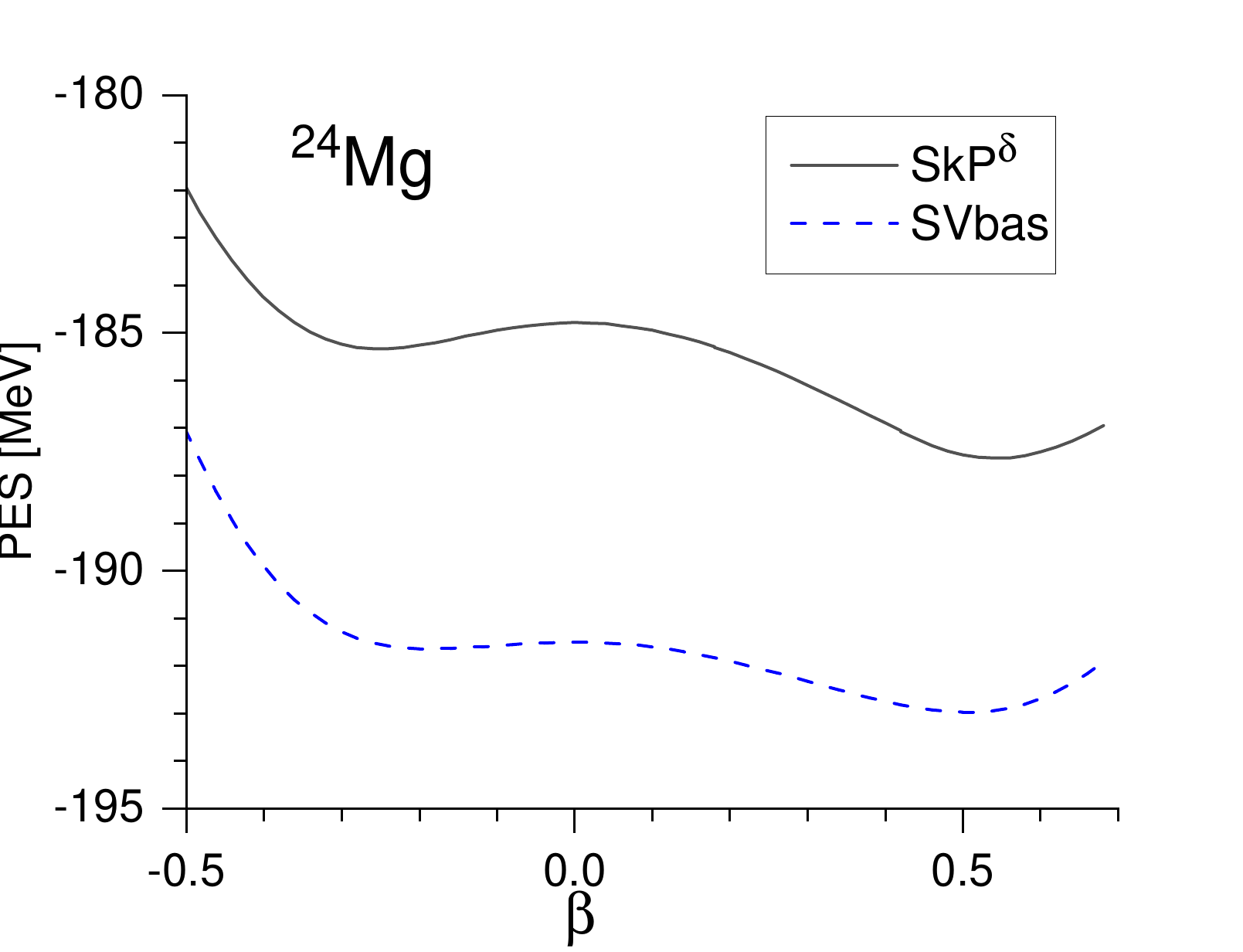}
 \caption{The calculated potential energy surface (PES) as a function of the axial quadrupole deformation $\beta$.}
 \label{fig1_PES}
 \end{center}
 \end{figure}

The single-particle spectra and pairing characteristics are calculated with the code \textsc{Skyax}~\cite{Skyax} on a two-dimensional (2D) grid in cylindrical coordinates.~A grid step of $0.4$~fm and a calculation box extending up to three nuclear radii were employed.~All proton and neutron single-particle (s-p) levels from the bottom of the potential well up to $+30$~MeV are included.~The QRPA calculations use a large two-quasiparticle (2qp) configuration space, ensuring that the isoscalar energy-weighted sum rule EWSR(IS$0$)~\cite{Harakeh} was exhausted by $97$–$99$\%.

The equilibrium axial quadrupole deformation $\beta$ is obtained at the minimum of the total nuclear energy as shown in Table~\ref{tab2:def+BE}.~The calculated deformation parameters are in reasonable agreement with the experimental value~\cite{nndc}.~Figure~\ref{fig1_PES} displays the potential energy surface (PES), which exhibits shallow prolate minima at $\beta$$=$$0.545$ (SkP$^{\delta}$) and $\beta$ $=$$0.527$ (SVbas), reflecting the deformation softness of $^{24}$Mg.~The calculated binding energies per nucleon $BE/A$ are slightly underestimated compared to experiment.

Pairing correlations are treated within the Bardeen-Cooper-Schrieffer (BCS) scheme~\cite{Repko_EPJA17_pairing}.~To deal with the divergent nature of the zero-range pairing interaction, an energy-dependent cut-off is applied~\cite{Repko_EPJA17_pairing,Be00}.~The spurious particle-number admixtures induced by pairing were removed using the method described in Ref.~\cite{Kvasil_EPJA19}.

The distribution of the isoscalar monopole (IS$0$, $\lambda$$=$$0$) and quadrupole (IS$20$, $\lambda\mu$$=$$20$) strengths is described by the strength functions\\
\begin{eqnarray}
\label{func_IS0}
S(\text{IS}0,E) &=& \sum_{\nu} \left| \langle \nu | \hat{M}_{\text{IS}0} | 0 \rangle \right|^2 \, \xi_{\Delta}(E - E_{\nu})~, \\[4pt]
\label{func_IS20}
S(\text{IS}20,E) &=& \sum_{\nu} \left| \langle \nu | \hat{M}_{\text{IS}20} | 0 \rangle \right|^2 \, \xi_{\Delta}(E - E_{\nu})~,
\end{eqnarray}\\
where $\nu$ denotes the QRPA (or unperturbed 2qp) excited state with energy $E_{\nu}$, and $|0\rangle$ is the ground QRPA (or 2qp) state.

The isoscalar monopole and quadrupole transition operators are defined as
\begin{equation}
\hat{M}_{\text{IS}0} = \sum_{i=1}^{A} r_i^2 Y_{00}(\hat{r}_i)~, 
\qquad
\hat{M}_{\text{IS}20} = \sum_{i=1}^{A} r_i^2 Y_{20}(\hat{r}_i)~,
\end{equation}
where $A$ is the mass number and $Y_{00}$, $Y_{20}$ are spherical harmonics.~For convenient comparison with experimental data, the discrete QRPA strength is folded with a Lorentzian function
\begin{equation}
\xi_{\Delta}(E - E_{\nu}) = 
\frac{\Delta}{2\pi\left[(E - E_{\nu})^2 + \Delta^2/4\right]}~.
\end{equation}
For the wavelet analysis, we use an averaging parameter of $\Delta = 70$~keV (corresponding to the experimental energy resolution) and an energy grid step of $dE = 10$~keV.

\begin{figure} 
 \begin{center}
  \includegraphics[scale=0.32]{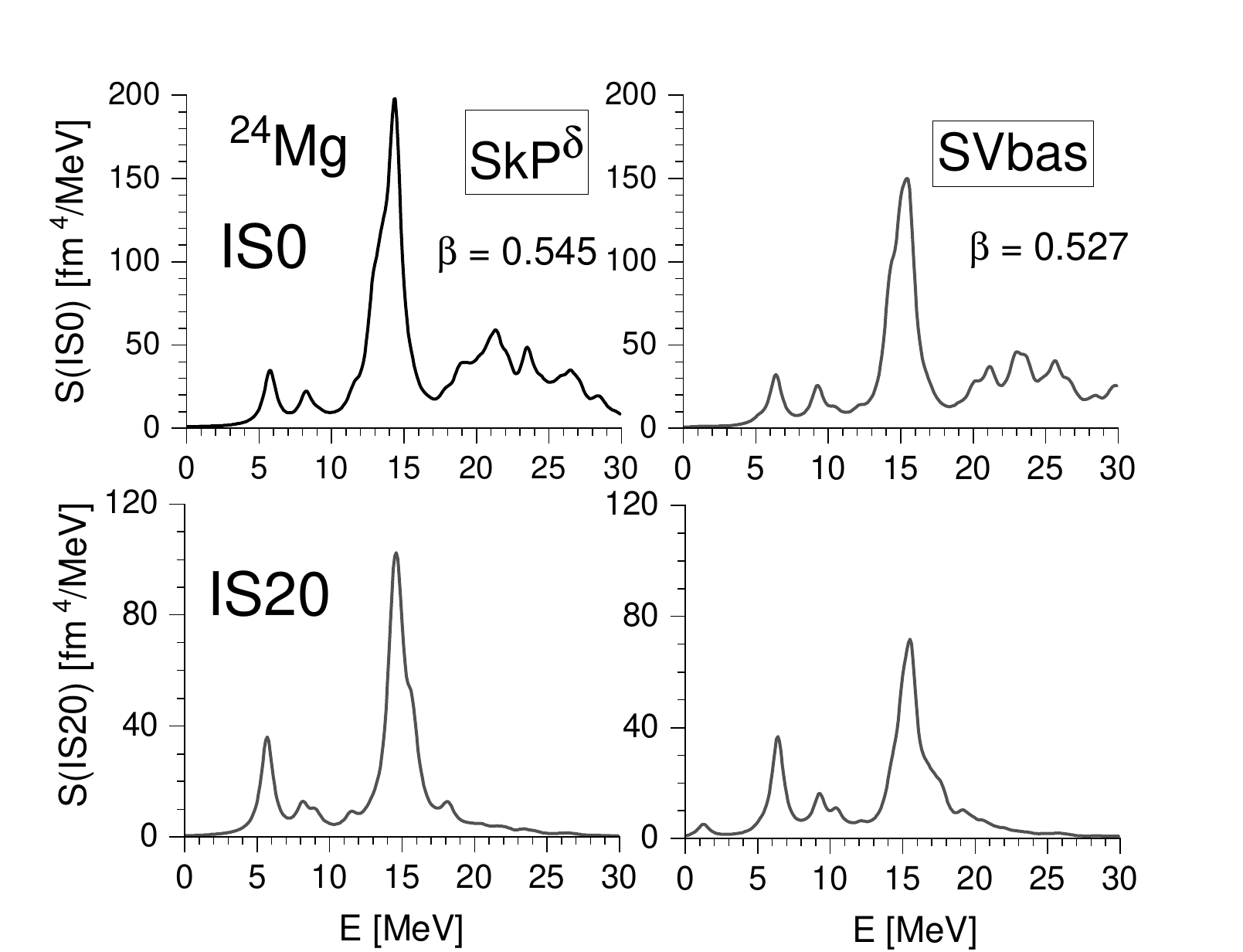}
 \caption{The monopole (top) and quadrupole (bottom) strength functions calculated with  SkP$^\delta$  (left) and SVbas (right) forces.}
 \label{fig2_IS0_IS20}
 \end{center}
 \end{figure}

\section{Monopole-quadrupole coupling} 
\label{sec4}

In Fig.~\ref{fig2_IS0_IS20}, the calculated isoscalar monopole (Eq.~(\ref{func_IS0})) and quadrupole (Eq.~(\ref{func_IS20})) strength functions in $^{24}$Mg are shown.~For convenient comparison, an averaging parameter of $\Delta$$=$$1$~MeV is used.~A pronounced correspondence between the IS$0$ and IS$20$ responses is observed in the energy range of $5$$-$$20$~MeV.~This feature originates from the deformation-induced MQC effect, discussed in detail in Ref.~\cite{Bahini_PRC22}.~As demonstrated in Ref.~\cite{Kva_PRC16} within the self-consistent separable QRPA model~\cite{Nest_PRC06}, the absence of MQC leads to the disappearance of the strong IS$0$ peaks below the ISGMR region.~Therefore, in $^{24}$Mg, the IS$0$ strength observed between $10$ and $18$~MeV can be mainly attributed to MQC, while the higher-energy part ($18$$-$$24$~MeV) is dominated by the ISGMR.

\section{Wavelet analysis}
\label{sec5}
Wavelet analysis  is a mathematical method widely used in various signal processing applications~\cite{Tor_wavelet98,Bha_wavelet20}.~In particular, wavelet methods can be implemented  for the analysis of spatial fields with a complex multiscale structure or temporal signals with a time-dependent spectral decomposition.~The wavelet transform  can be roughly considered as a local Fourier transform.~In nuclear physics, the Continuous Wavelet Transform (CWT) is mainly used for the extraction of characteristic scales $\delta E$ which are associated with nuclear decay properties in a local energy interval of nuclear excitations~\cite{Shev_PRC08,Shev_PRC09,Bahini_PRC24}.~The choice of the wavelet form (mother wavelet) plays an important role.~In nuclear physics, the Morlet mother wavelet (a product of the complex sinusoidal function and Gaussian function) is generally used~\cite{Shev_PRC08,Shev_PRC09,Kureba_PLB18,Bahini_PRC24}.~This choice is explained by the fact that detector responses in nuclear experiments are usually approximated by a Gaussian line shape.~In this study, we use the complex-Morlet mother wavelet\\
\begin{equation}
	\label{e271a}
	\Psi(x) = \dfrac{1}{\pi^{\frac{1}{2}}f_\text{b}}\exp\left(2\pi if_\text{c}x - \frac{x^2}{f_\text{b}}\right)~,
\end{equation}\\	
where $f_\text{b}$$=$$2$ and $f_\text{c}$$=$$1$ correspond to the wavelet bandwidth and the center frequency of the wavelet, respectively.~The convolution of a given signal $\sigma(E)$ with the wavelet function (generally complex-conjugated) yields the coefficients of the wavelet transform\\
\begin{equation}
\label{coef}
C(\delta E, E_\text{x})=\frac{1}{\sqrt{\delta E}}\int \sigma(E)\Psi^* \Bigg( \frac{E_\text{x}-E}{\delta E} \Bigg) dE~,
\end{equation}\\
where the scaling factor $\delta E$ scales (expands or suppresses) the mother wavelet  while  $E_\text{x}$ shifts the wavelet position along the excitation-energy range and thus changes the scale localization.~The two-dimensional transform $C(\delta E, E_\text{x})$ is large if the mother wavelet matches well the signal $\sigma(E)$ at the localization $E_\text{x}$.

The extraction of wavelet energy scales can be achieved from the wavelet coefficient plot as peaks in the corresponding power spectrum\\
\begin{equation}
	\label{power}
P(\delta E) = \dfrac{1}{N}\sum_{i = 1}^{N} C_i(\delta E) \cdot C_i^*(\delta E)~,
\end{equation}\\
as a function of energy scale $\delta E$.~The sum over the index $i =1,\cdots,N$ with $N$ being the number of energy bins in the excitation-energy region is considered.~In the present study, we use $1600$ bins (of $10$ keV each) in the range $9$$-$$25$ MeV.~The existence of distinctive scales is indicated by the local maxima and inflection points within the power spectrum.

In this study, we employ two complementary representations of the wavelet transform: the real part, \( \mathrm{Re}[C(\delta E, E_\text{x})] \), and the modulus, $|C(\delta E, E_\mathrm{x})|$$=$$\sqrt{[\mathrm{Re}(C)]^2 + [\mathrm{Im}(C)]^2}$.~The real part has been commonly used in previous nuclear wavelet analyses (see, e.g., Ref.~\cite{Bahini_PRC24} and references therein).~Because the real and imaginary components of the wavelet coefficients have similar local magnitudes and carry complementary information on the local phase and symmetry of the strength distribution, the real part alone can capture the characteristic energy scales.~However, its oscillatory nature can complicate a visual interpretation.~The modulus, on the other hand, provides a phase-independent and smoother representation of the energy localization.~Both representations yield the same wavelet powers and characteristic scales, and are therefore equally suitable for the analysis of the ISGMR fine structure in $^{24}$Mg.

\begin{figure*}[t] 
 \begin{center}
\includegraphics[width=0.95\textwidth]{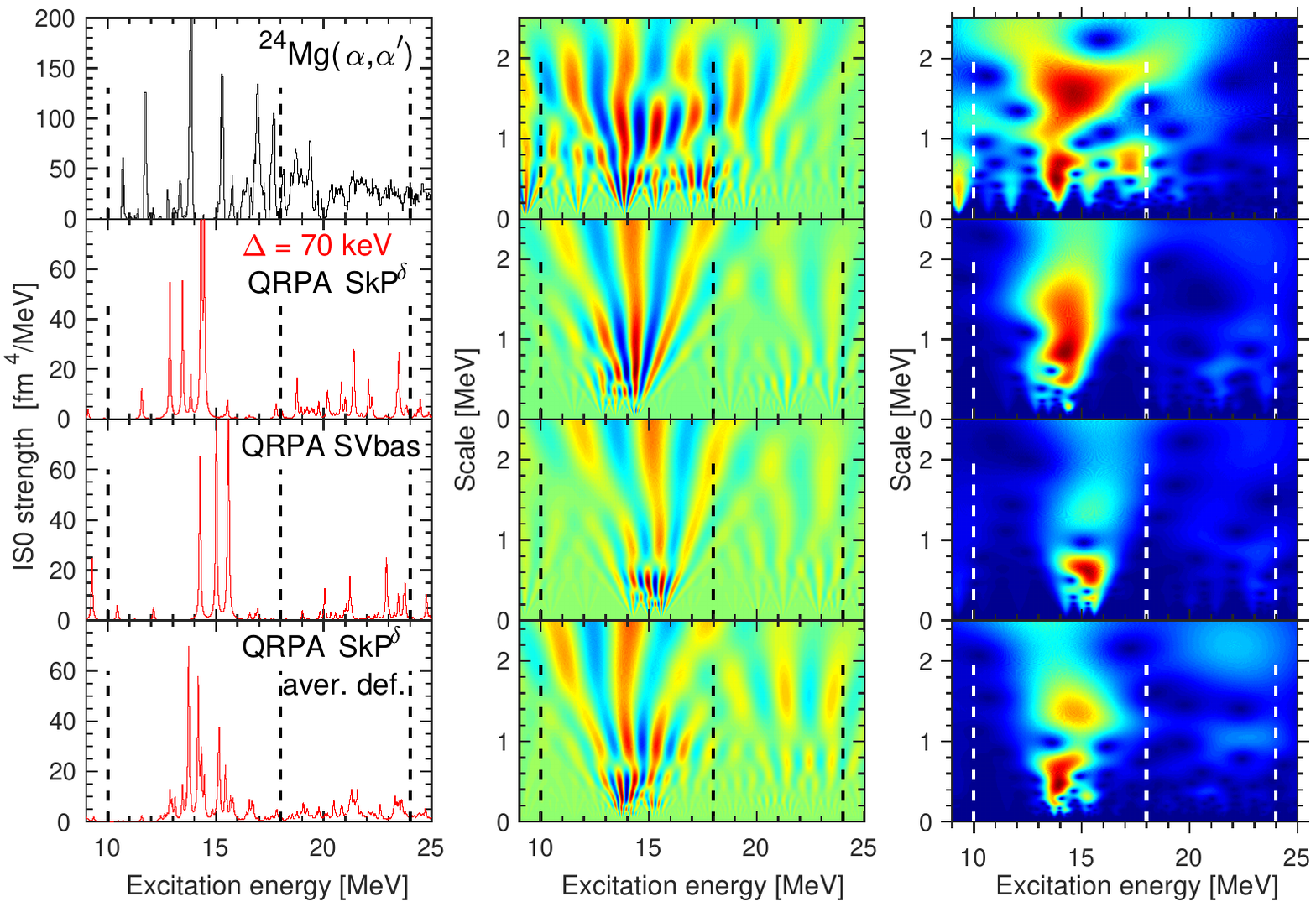}
  \caption{
Left column: Experimental isoscalar monopole strength distribution in \(^{24}\)Mg compared with QRPA model predictions folded with the experimental energy resolution ($\Delta E$$=$$70~\text{keV}$).~Middle column: Real parts of the corresponding wavelet transforms, \(\mathrm{Re}[C(\delta E, E_\text{x})]\).~Right column: Moduli of the wavelet transforms, \(|C(\delta E, E_\text{x})|\).~Vertical dashed lines indicate the excitation-energy intervals ($10$$-$$24$ MeV, $10$$-$$18$ MeV and $18$$-$$24$ MeV) over which the wavelet coefficients are summed to obtain the power spectra displayed in Fig.~\ref{powers}.
}
\label{fig3:strengths+WL}
 \end{center}
 \end{figure*}
\begin{figure*}[t] 
 \begin{center}
\includegraphics[width=0.95\textwidth]{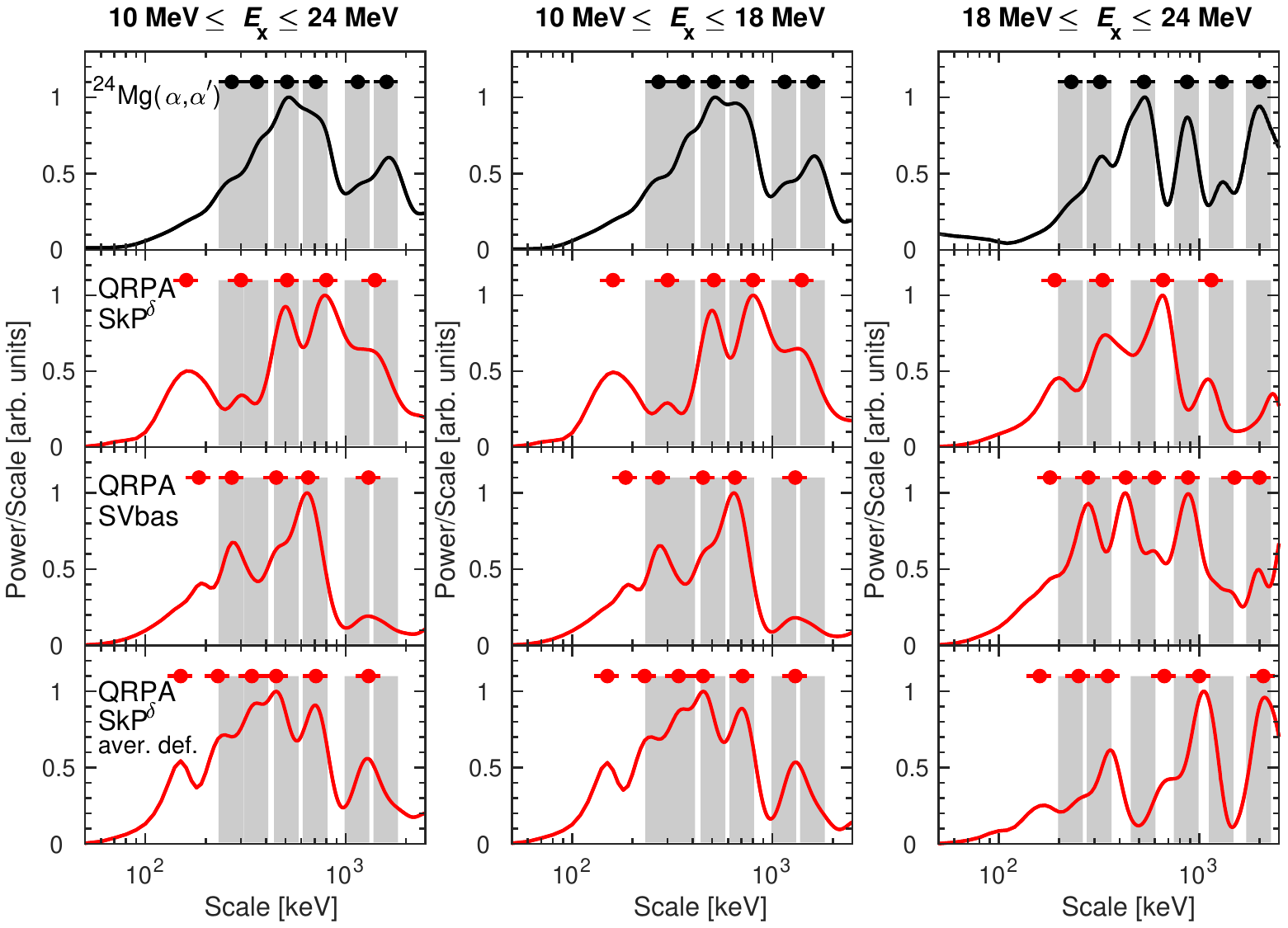}
  \caption{Experimental (top) and theoretical QRPA power spectra calculated with the SkP$^{\delta}$ (middle, including the deformation-averaged strength function shown in the lowest row) and SVbas (bottom) interactions for $^{24}$Mg.~Results are shown for the total excitation-energy range $10$$–$$24$~MeV (left), the MQC region $10$$–$$18$~MeV (middle), and the ISGMR region $18$$–$$24$~MeV (right).~Energy scales and their associated errors are indicated by filled circles and horizontal bars, respectively.~In the experimental panels (top), the additional vertical gray bars denote the $1\sigma$ statistical errors; the same gray bars are reproduced in the theoretical panels for direct comparison.}
\label{powers}

  \label{powers}
 \end{center}
 \end{figure*}

\section{Results and discussion}
\label{sec6}
\subsection{Strength distributions and wavelet transforms}

In the left column of Fig.~\ref{fig3:strengths+WL}, the experimental IS$0$ strength distribution is compared with the calculated SkP$^{\delta}$ and SVbas strength functions (Eq.~(\ref{func_IS0})) for $^{24}$Mg.~Both SkP$^{\delta}$ (second row of  Fig.~\ref{fig3:strengths+WL}) and SVbas (third row of  Fig.~\ref{fig3:strengths+WL}) generally reproduce the prominent IS$0$ peaks observed between $10$ and $16$~MeV.~Following the discussion of Fig.~\ref{fig2_IS0_IS20} and the detailed analysis in Ref.~\cite{Bahini_PRC22}, these peaks originate from the deformation-induced coupling between monopole and quadrupole $K$$=$$0$ excitations, the MQC effect.~The considerable magnitude of these peaks reflects the large quadrupole deformation of $^{24}$Mg: the stronger the deformation, the more pronounced the MQC effect~\cite{Kva_PRC16}.

As seen in Fig.~\ref{fig3:strengths+WL}, neither SkP$^{\delta}$ nor SVbas reproduces the high IS$0$ peaks measured at $16$$–$$18$~MeV.~Furthermore, in the ISGMR region ($18$$–$$24$~MeV), the calculated strength is weaker than the experimental one, particularly for SVbas.~It should be emphasized that our calculations do not miss any IS$0$ strength since the EWSR(IS$0$) is almost fully exhausted.~The apparent deficit of local strength is therefore more likely due to differences in the energy distribution of experimental and theoretical IS$0$ strengths.~For spherical nuclei, it has been shown that coupling to complex configurations (CCC), which is neglected in the present QRPA calculations, can shift a portion of the high-energy IS$0$ strength~\cite{coloprl2023} down toward the ISGMR region and below.~Including such CCC effects could thus improve the agreement with experiment in our case.~Moreover, as discussed later, the agreement can also be enhanced by accounting for the deformation softness of $^{24}$Mg.

The middle and right panels of Fig.~\ref{fig3:strengths+WL} show the corresponding wavelet analyses using two complementary representations: $\mathrm{Re}[C(\delta E, E_{\mathrm{x}})]$ and $|C(\delta E, E_{\mathrm{x}})|$.~Both quantities depend on the excitation energy $E_{\mathrm{x}}$ and the wavelet scale $\delta E$.~In the oscillatory maps of $\mathrm{Re}[C(\delta E, E_{\mathrm{x}})]$ (middle panels), alternating blue and red regions indicate negative and positive values, respectively, reflecting the phase structure of the wavelet transform.~In contrast, the $|C(\delta E, E_{\mathrm{x}})|$ maps (right panels) display the magnitude of the coefficients: blue regions correspond to low absolute values, whereas red regions indicate high absolute values.

Figure~\ref{fig3:strengths+WL} shows that both wavelet presentations, \( \mathrm{Re}[C(\delta E, E_\text{x})] \) and $|C(\delta E, E_\text{x})|$, clearly correspond to the distributions of IS$0$ strength in the left panels.~Namely, the characteristic energy scales are mainly concentrated at $10$$-$$20$ MeV, which corresponds to the region of the pronounced IS$0$ peaks in the strength functions.~In this region, the experiment shows the maximal densities for the scales $\delta E$$\sim$$200$$-$$800$ and $1000$$-$$2000$ keV.~At the equilibrium deformation, SkP$^{\delta}$ yields maximal scale densities for $\delta E$$\sim$400$-$1800~keV, in reasonable agreement with the experiment.~SVbas produces a bright density spot around $\delta E$$\sim$200$-$$800$~keV; however, unlike the data, the following structure at $1$$-$$2$~MeV appears significantly weaker.~Both SkP$^{\delta}$ and SVbas show a low density of scales at $18$$-$$20$~MeV, consistent with the drop in the IS0 strength observed in the corresponding strength distributions.~This reduction is visible in the right panels as a blue pattern.~As seen in Fig.~\ref{fig3:strengths+WL}, the two wavelet representations, \( \mathrm{Re}[C(\delta E, E_\text{x})] \) and \( |C(\delta E, E_\text{x})| \), provide consistent results.

\subsection{Wavelet powers and scales}

The middle and right panels of Fig.~\ref{fig3:strengths+WL} demonstrate rather smooth distribution of the scales.~Nevertheless, it would be informative to perform a more detailed analysis and extract particular dominant scales, as was done in numerous previous studies of the nuclear wavelet, see e.g.~\cite{Bahini_PRC24} and references therein.~The dominant scales can be estimated through the power $P(\delta E)$ of the wavelet transforms.~In Fig.~\ref{powers}, the experimental and theoretical power spectra (Eq.~(\ref{power})) are displayed for the energy intervals of interest.~The dominant energy scales are displayed as black (experiment) and red (theory) filled circles.~The associated error is given by one standard deviation of the corresponding width-like scale at half of the peak width (FWHM), cf.~\cite{Bahini_PRC24}.~In each panel, the powers are normalized by their maximal value within the given energy range.~The top-left panel of Fig.~\ref{powers} shows that, in the excitation energy region $10$$-$$24$ MeV, the experimental strength is dominated by scales of $\delta E$$=$$200$$-$$1000$ keV.~Scales of $\delta E = 1000$$-$$2000$ keV, although producing a noticeable peak in the power, are less pronounced.~The broad power hump in the range $\delta E$$=$$200$$-$$1000$ keV exhibits several local maxima and inflections, indicated by black dots.~These local maxima are the characteristic energy scales $\delta E$.~For comparison purposes and in order to facilitate the determination of similar scales in the corresponding power spectra from the model calculations, the results obtained from experiments are also displayed as vertical grey bars in all panels of Fig.~\ref{powers}.~For the sake of better display, their widths have been reduced to $2/3$ of the standard deviation.

\begin{table} 
	\caption{Dominant scales extracted in the whole excitation energy region $10~\text{MeV}$$\leq E_{\text{x}}$$ \leq $$~24~\text{MeV}$.~Equivalent characteristic energy-scale values are vertically aligned.}	
	\label{tab3:dE_10-24}
	\begin{center}
		\setlength{\arrayrulewidth}{0.5pt}
		\setlength{\tabcolsep}{0.3cm}
		\renewcommand{\arraystretch}{1.5}	
		\begin{tabular}{cccccccccc}
			\hline\hline
			\multicolumn{10}{l}{Dataset \hspace{3cm}Scales (keV)\hspace{1.5cm}~~}\\
			\hline
			Expt.~&  & $270$ $360$ & $510$ & $710$ & $1150$ $1600$ &
			\\
			SkP$^\delta$ & $160$ & $300$ & $510$ & $800$ & $1400$ &
			\\
  		    SVbas & $185$ & $270$ & $450$& $650$ & $1300$ &
			\\
			\hline\hline	
		\end{tabular}
	\end{center}	
\end{table}
\begin{table} 
	\caption{Dominant MQC scales extracted in the excitation energy region $10~\text{MeV}$$\leq E_{\text{x}}$$ \leq $$~18~\text{MeV}$.~Equivalent characteristic energy-scale values are vertically aligned.}	
	\label{tab4:dE_10-18}
	\begin{center}
		\setlength{\arrayrulewidth}{0.5pt}
		\setlength{\tabcolsep}{0.3cm}
		\renewcommand{\arraystretch}{1.5}	
		\begin{tabular}{cccccccccc}
			\hline\hline
			\multicolumn{10}{l}{Dataset \hspace{3cm}Scales (keV)\hspace{1.5cm}~~}\\
			\hline
			Expt.~&  & $270$ $360$ & $490$ & $710$ & $1150$ $1600$ &
			\\
			SkP$^\delta$ & $160$ & $300$ & $510$ & $800$ & $1400$ &
			\\
  		    SVbas & $185$ & $270$ & $450$& $650$ & $1300$ &
			\\
			\hline\hline	
		\end{tabular}
	\end{center}	
\end{table}

\begin{table} 
	\caption{Dominant ISGMR energy scales extracted in the excitation energy region $18~\text{MeV} \leq E_{\text{x}} \leq 24~\text{MeV}$.~Equivalent characteristic energy-scale values are vertically aligned.}	
	\label{tab5:dE_18-24}
	\begin{center}
		\setlength{\arrayrulewidth}{0.5pt}
		\setlength{\tabcolsep}{0.22cm}
		\renewcommand{\arraystretch}{1.5}	
		\begin{tabular}{ccccccccccc}
			\hline\hline
			\multicolumn{10}{l}{Dataset \hspace{2.25cm}Scales (keV)\hspace{2.25cm}~~}\\
			\hline
			Expt.~& $230$ & $320$ & & $530$ & $870$ & $1300$ & $2000$ &
			\\
			SkP$^\delta$ & $200$ & $330$ &  & $660$ &  & $1150$ & &
			\\
  		    SVbas & $180$ & $280$ & $430$ & $600$ & $870$ & $1500$ & $2000$ &
			\\
			\hline\hline	
		\end{tabular}
	\end{center}	
\end{table}
 \begin{figure} 
 \begin{center}
  \includegraphics[scale=0.32]{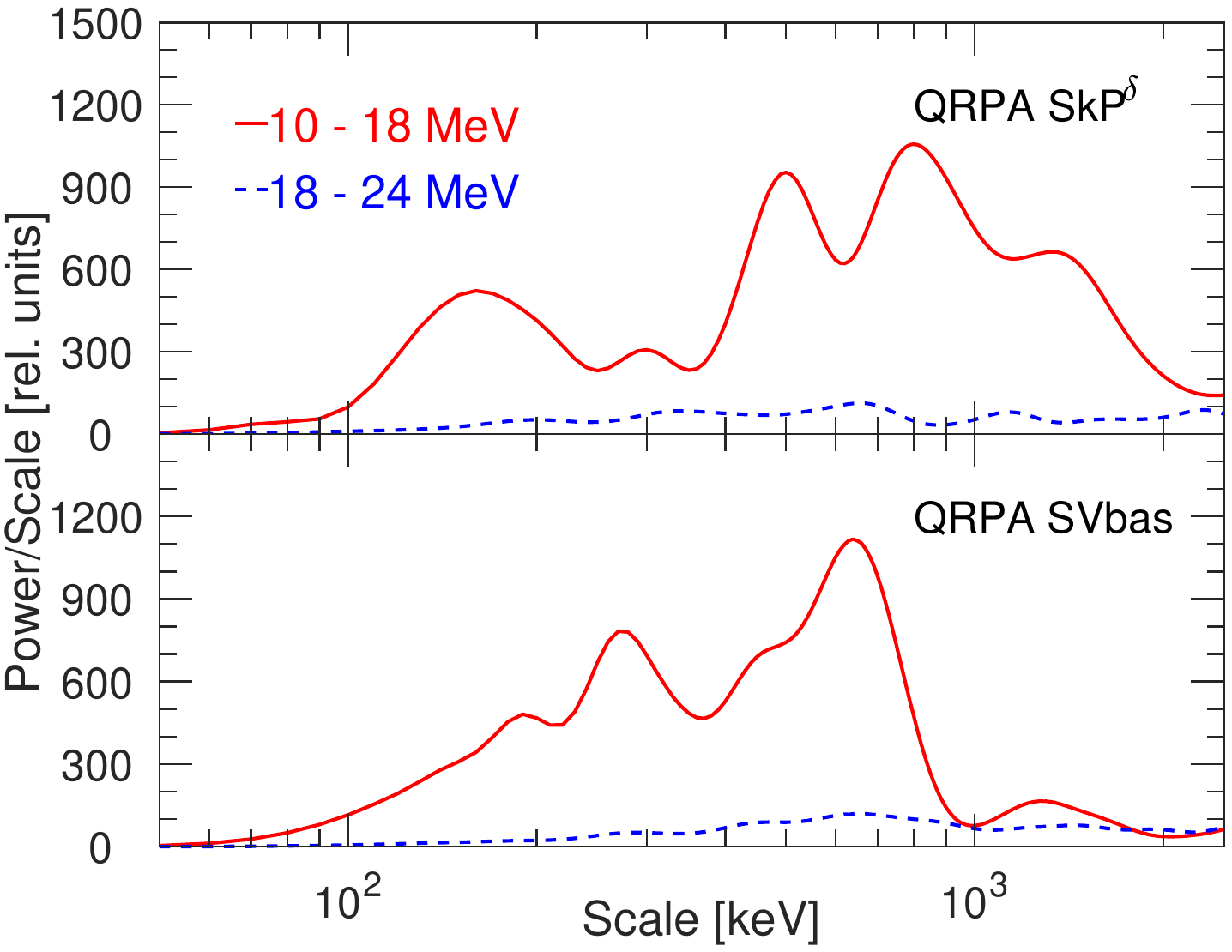}
 \caption{Non-normalized SkP$^\delta$ (top) and SVbas (bottom) powers for MQC ($10$$-$$18$ MeV) and ISGMR ($18$$-$$24$ MeV) energy ranges.}
 \label{S0_10_16_vs_16_25}
 \end{center}
 \end{figure}
 The SkP$^{\delta}$ calculations qualitatively reproduce this broad power hump at $\delta E$$=$$200$$-$$1000$ keV but the local maxima are more pronounced.~In addition, SkP$^{\delta}$ predicts a peak at $\delta E$$=$$160$ keV, which is absent in the experiment.~This feature likely arises from a local drop in the power at $\delta E$$=$$100$$-$$300$ keV, possibly due to neglecting CCC effects and the deformation softness of $^{24}$Mg.~The SVbas calculations exhibit distinct humps at $\delta E$$\sim$$300$ and $660$ keV, resulting in a power profile that deviates more noticeably from the experimental distribution.~The experimental and theoretical characteristic scales are summarized in Table~\ref{tab3:dE_10-24}.~Following previous studies~\cite{Shev_PRC08,Shev_PRC09}, the scales are categorized into three groups: small scales around $100$ keV, medium scales at $200$$-$$1000$ keV, and large scales of the order of several MeV.~Table~\ref{tab3:dE_10-24} indicates that, at least for the medium scales, the theoretical results are in reasonable agreement with the experimental data.

\begin{figure*} 
 \begin{center}
 \includegraphics[width=0.95\textwidth]
 {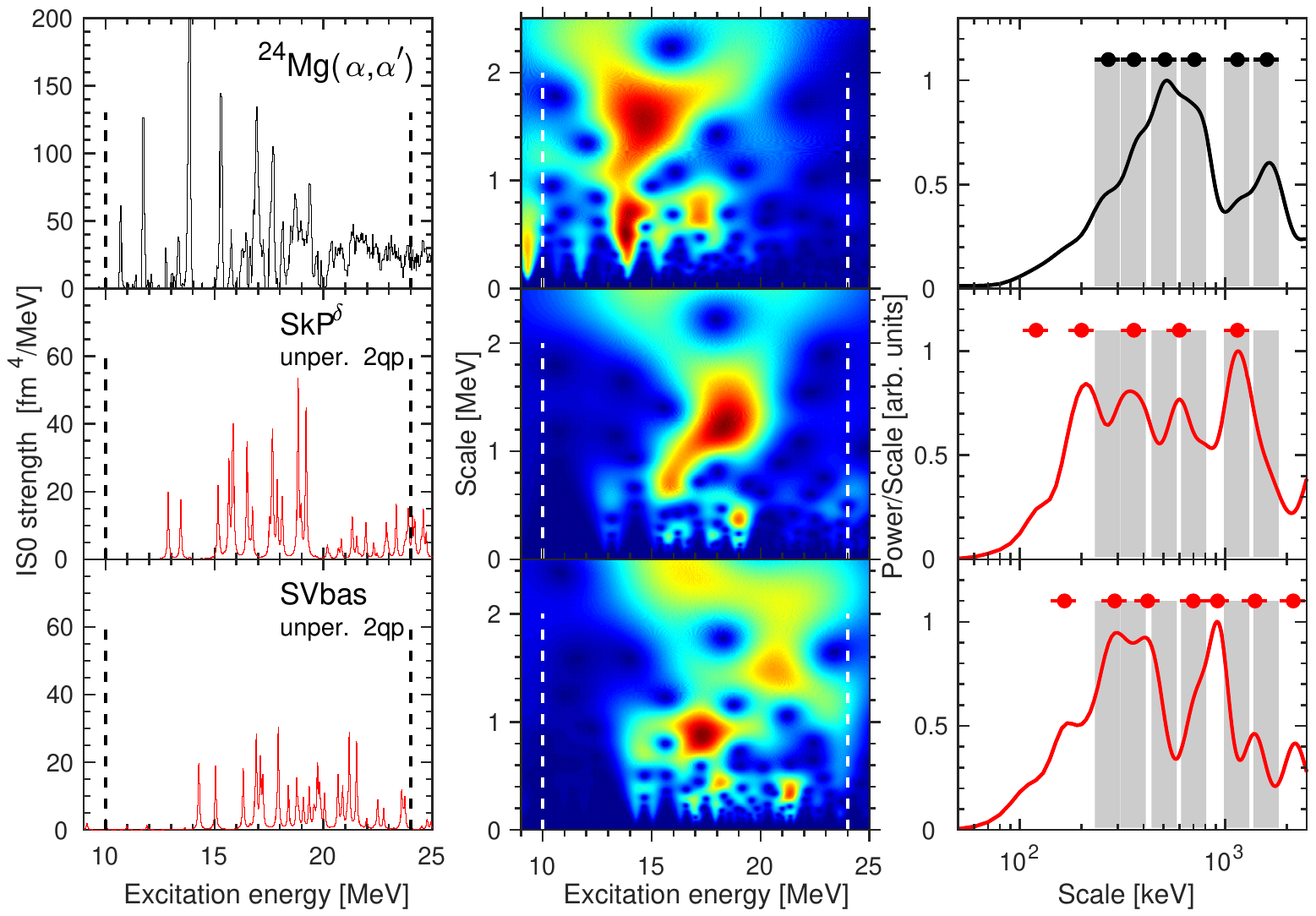}
 \caption{Left column: Experimental isoscalar monopole strength distribution in \(^{24}\)Mg compared with unperturbed $2$qp model predictions folded with the experimental energy resolution (\(\Delta E = 70~\text{keV}\)).~Middle column: Moduli of the corresponding wavelet transforms, \(|C(\delta E, E_\text{x})|\).~Vertical dashed lines in the first two columns indicate the excitation-energy interval ($10$$-$$24$~MeV) over which the wavelet coefficients are summed to obtain the power spectra shown in the right column.}
 \label{S0-2qp}
 \end{center}
 \end{figure*}

As discussed above, the IS$0$ states in the regions $10$$–$$18$ and $18$$–$$24$~MeV have different physical origins~\cite{Bahini_PRC22}: the former arise primarily from deformation-induced MQC, whereas the latter correspond to the ISGMR.~It is, therefore, insightful to perform the wavelet analysis separately for these two energy intervals.~The middle column of Fig.~\ref{powers} and Table~\ref{tab4:dE_10-18} present the wavelet results obtained for the MQC region.~We observe that both the experimental and theoretical power spectra are similar to those obtained for the total excitation-energy range $10$$-$$24$~MeV.~This indicates that the IS$0$ strength in the MQC region ($10$$-$$18$~MeV) predominantly determines the overall wavelet response across the full energy range.~For the ISGMR region ($18$$-$$24$~MeV), the calculated normalized powers in the right column of Fig.~\ref{powers} and the characteristic scales in Table~\ref{tab5:dE_18-24} are in acceptable agreement with the experiment.

 The powers discussed in the previous sections were normalized to the maximum value in each wavelet power spectrum.~To compare the {\it relative} contributions of MQC and ISGMR components, it is useful to examine their {\it non-normalized} power values.~This comparison is presented in Fig.~\ref{S0_10_16_vs_16_25}, where the relative powers for SkP$^\delta$ and SVbas are shown for the MQC and ISGMR energy ranges.~It is evident that, in the SkP$^\delta$ case, almost all characteristic scales originate predominantly from the MQC region.~For SVbas, the same behavior is observed for $\delta E$$<$$1000$~keV.~Therefore, in agreement with the previous discussion, the MQC is mainly responsible for the fine structure in the $10$$-$$24$~MeV region of $^{24}$Mg.

\subsection{Effect of the deformation softness}

As shown in Fig.~\ref{fig1_PES}, the prolate minimum in the PES of $^{24}$Mg is rather shallow, indicating that this nucleus exhibits a noticeable softness with respect to axial quadrupole deformation.~Consequently, excited monopole states may occur at quadrupole deformations deviating from the equilibrium values, $\beta$$=$$0.545$ for SkP$^\delta$ and $\beta$$=$$0.527$ for SVbas.~To assess the impact of this deformation softness, we performed deformation-constrained SkP$^\delta$ calculations for a set of deformations around the equilibrium value $\beta$$=$$0.545$ and computed the averaged strength function\\
\begin{equation}
S_{\rm aver}(\text{IS0},E) = \frac{1}{5}\sum_{i=1}^{5} S(\text{IS0},E,\beta_i),
\end{equation}\\
with $\beta_1$$=$$0.45$, $\beta_2$$=$$0.50$, $\beta_3$$=$$0.545$, $\beta_4$$=$$0.60$, and $\beta_5$$=$$0.65$.~The results are shown in the bottom panels of Figs.~\ref{fig3:strengths+WL} and~\ref{powers}.~We see that the averaged strength noticeably improves the  agreement with the experimental data.~In Fig.~\ref{fig3:strengths+WL}, the SkP$^\delta$ calculations now reproduce much better the experimental distribution of the scale density for $\delta E$$=$$200$$-$$1000$ keV.~In Fig.~\ref{powers}, the artificial hump at $\delta E$$=$$160$~keV in the calculated power spectrum is now partly suppressed.~These results indicate that deformation softness plays an essential role in describing the wavelet characteristics of $^{24}$Mg.~Furthermore, the calculations based on $S_{\rm aver}(\text{IS0},E)$ confirm that the total power distribution in the range $10$$-$$24$~MeV is almost entirely determined by the MQC region ($10$$-$$18$~MeV).

\subsection{Landau damping}

\begin{figure} 
\begin{center}
\includegraphics[scale=0.32]{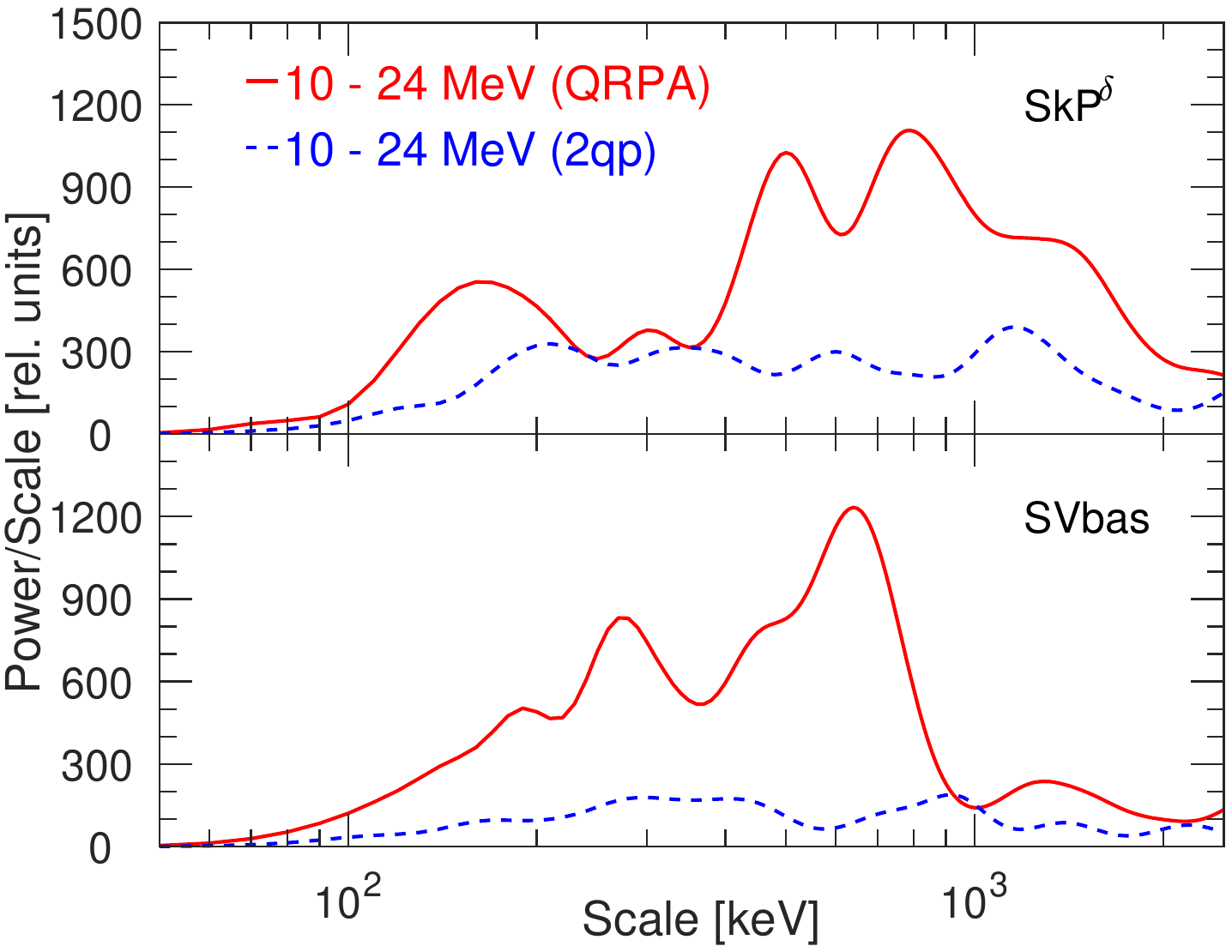}
\caption{SkP$^\delta$ and SVbas non-normalized powers in the energy range $10$$–$$24$ MeV, obtained with (QRPA) and without (2qp) residual interaction.}
 \label{powers:QRPA-2qp}
\end{center}
 \end{figure}

To clarify the role of Landau damping, it is instructive to compare the strength distributions and wavelet results obtained from QRPA and unperturbed two-quasiparticle ($2$qp) spectra.~The $2$qp results for SkP$^\delta$ and SVbas are compared with the experimental data in Fig.~\ref{S0-2qp}.~It is evident that the calculated $2$qp strength distributions, wavelet amplitudes $|C(\delta E, E_\text{x})|$, and power spectra differ substantially from the experimental patterns.~Hence, a pure $2$qp description is clearly insufficient, and the inclusion of QRPA correlations is essential.~A comparison of the $2$qp strength functions in Fig.~\ref{S0-2qp} with the QRPA results in Fig.~\ref{fig3:strengths+WL} shows that, relative to QRPA, the $2$qp monopole strength is enhanced in the intermediate region ($16$$-$$20$~MeV) and, conversely, suppressed in the MQC range.~The latter behavior can be explained by the fact that, in the absence of the residual interaction, the ISGQR ($K$$=$$0$) peak lies above the $10$$-$$15$~MeV region, resulting in a weak MQC contribution there (see the detailed discussion in Ref.~\cite{Bahini_PRC22}).~Thus, it is precisely the strong isoscalar monopole and quadrupole residual interactions and, hence, the associated Landau damping that give rise to the actual IS$0$ strength distributions displayed in Fig.~\ref{fig3:strengths+WL} and to the pronounced MQC effect.

This conclusion is further supported by Fig.~\ref{powers:QRPA-2qp}, which compares the relative $2$qp and QRPA power distributions.~It is seen that the QRPA contribution dominates over that of the $2$qp.~Therefore, the residual interaction is crucial not only for shaping the IS$0$ strength distribution but also for generating the large integrated weight of the characteristic scales.~These findings once again underscore the essential role of Landau damping in monopole excitations of deformed nuclei.

\section{Conclusions}
\label{sec7}
The wavelet analysis of the iThemba LABS $(\alpha,\alpha')$ experimental data for monopole excitations in $^{24}$Mg is performed within a fully self-consistent quasiparticle random-phase approximation (QRPA) framework employing the Skyrme parametrizations SkP$^{\delta}$ and SVbas.~Two types of isoscalar monopole (IS$0$) excitations of different physical origin were considered: (i) the IS$0$ states in the region $10$--$18$~MeV, arising from the deformation-induced coupling (MQC) between monopole and quadrupole ($K$$=$$0$) modes, and (ii) the main isoscalar giant monopole resonance (ISGMR) located at higher excitation energies.

Overall, a reasonable description of the IS$0$ strength functions and the associated wavelet characteristics (wavelet transforms and powers) is achieved, particularly with the SkP$^{\delta}$ interaction, which features a low nuclear matter incompressibility $K_{\infty}$$=$$202$~MeV.~From the wavelet power spectra, the characteristic scales $\delta E$ associated with Landau dissipation widths were extracted.~However, the IS$0$ excitations in the strongly deformed nucleus $^{24}$Mg are characterized by smooth distributions in the range $\delta E$$=$$200$$-$$1000$ and $1000$$-$$2000$~keV, rather than by discrete, well-separated scales.~Furthermore, the analysis demonstrates that, over the total excitation energy interval $10$$-$$24$~MeV, the MQC contribution to the wavelet powers strongly dominates over that of the ISGMR.~It is worth noting that, in the case of a dense spectrum with overlapping peaks, which partially occurs for monopole excitations in $^{24}$Mg, the spacings between the peaks also influence the wavelet characteristics.~Hence, one should exercise caution in directly associating the scales $\delta E$ with particular decay widths.

A comparison between the QRPA and unperturbed two-quasiparticle ($2$qp) results reveals the crucial role of the residual interaction in shaping both the strength distributions and the corresponding wavelet features.~Within QRPA, the weight of the dominant scales of $\delta E$$=$$200$$-$$1000$~keV is considerably enhanced relative to the $2$qp case.~Thus, Landau damping of one-phonon states manifests itself not only in the IS$0$ strength distribution but also in the increased weight of the dominant energy scales.

Despite the overall agreement, the calculated QRPA IS$0$ strength functions and wavelet powers in $^{24}$Mg still deviate in detail from the experimental results.~The agreement can be notably improved when the deformation softness of $^{24}$Mg is taken into account.~However, the remaining discrepancies indicate that this effect alone is insufficient and that the coupling of one-phonon QRPA states with more complex configurations—neglected in the present calculations—is likely to play a relevant role.

\section*{Acknowledgement}
The authors thank the Accelerator Group at iThemba LABS for the high-quality dispersion-matched beam provided for this experiment.~This work was supported by the National Research Foundation (NRF) of South Africa (Grant No.~$85509$, $118846$, $129603$ and REP SARC$180529336567$), as well as the BLTP JINR-SA grant for $2022$$-$$2024$.~A.B.~acknowledges financial support through iThemba LABS, the NRF of South Africa and the Laboratoire de Physique Corpusculaire de Caen (LPC Caen).~A.R.~acknowledges support by the Slovak Research and Development Agency under Contract No.~APVV-24-0516 and by the Slovak grant agency VEGA (Contract No.~2/0175/24).


\end{document}